\documentstyle[12pt]{article}
\def\appendix#1{
\addtocounter{section}{1}
\setcounter{equation}{0}
\renewcommand{\thesection}{\Alph{section}}
\section*{Appendix \thesection\protect\indent #1}
\addcontentsline{toc}{section}{Appendix \thesection\ \ \ #1}
}

\def\be{\begin{equation}}
\def\la{\label}
\def\ee{\end{equation}}
\def\bea{\begin{eqnarray}}
\def\eea{\end{eqnarray}}
\def\eps{\varepsilon}

\def\s{\sigma}
\def\n{\nabla}

\def\d{\delta}

\def\l{\left(}
\def\r{\right)}
\def\p{\partial}

\begin{document}
\title{
Antisymmetric tensor field on $AdS_5$.
}
\author{G.E.Arutyunov\thanks{arut@genesis.mi.ras.ru}
\mbox{} and S.A.Frolov
\thanks{
Address after September 1, 1998: The University of Alabama, Department
of Physics and Astronomy, Box 870324, Tuscaloosa, Alabama 35487-0324.
}
\mbox{} \\
\vspace{0.4cm}
Steklov Mathematical Institute,
\vspace{-0.5cm} \mbox{} \\
Gubkin str.8, GSP-1, 117966, Moscow, Russia;
\vspace{0.5cm} \mbox{}
}
\date {}
\maketitle
\begin{abstract}
By using the Hamiltonian version of the AdS/CFT
correspondence, we compute the two-point Green function of a local operator
in $D=4$ ${\cal N}=4$ super Yang-Mills theory, which corresponds to a massive
antisymmetric tensor field of the second rank on the $AdS_{5}$ background.
We discuss the conformal transformations induced on the boundary  by isometries
of $AdS_{5}$.
\end{abstract}

The recent Maldacena's conjecture \cite{M} relates the large $N$ limit
of certain  conformal theories in $d$-dimensions with classical
supergravity on the product of anti de Sitter space $AdS_{d+1}$ with a compact manifold.
According to \cite{GKP,W} the precise relation
consists in existing the correspondence between
supergravity fields and the set of local CFT operators.
Then the generating functional of the connected Green functions
of the CFT operators is identified with the on-shell
value of the supergravity action. With this identification at hand,
the AdS/CFT correspondence was recently tested by explicit computation
of some two- and three-point correlation functions of local operators in
$D=4$ ${\cal N}=4$ super Yang-Mills theory, which correspond to
scalar, vector, symmetric tensor and spinor fields
on the $AdS_5 $ background \cite{AV}-\cite{SSRS}.

$D=4$ ${\cal N}=4$ super Yang-Mills is related to the $S^5$
compactification of $D=10$ $IIB$ supergravity.
Except the fields mentioned above, the spectrum of
the compactified theory also contains the massive antisymmetric tensor
fields of the second rank \cite{KWN,GM}. These fields obey first-order
differential equations and their bulk action vanishes on shell.
Thus, the bulk action is not enough to compute the CFT Green functions
and one has to add some boundary terms. This is quite similar
to the case of fermions on the $AdS$ background \cite{HS}.
The origin of boundary terms in the AdS/CFT correspondence
was recently clarified in \cite{AF},
where it was shown that they appear in passing from the Hamiltonian
description of the bulk action to the Lagrangian one.
The idea was to treat the coordinate in the bulk direction
as the time and to present the bulk action in the form
$\int (p\dot{q}-H(p,q)+total~derivative)$. Here a choice of
coordinates and momenta is dictated by the transformation properties
of gravity fields under isometries of $AdS$. In the Hamiltonian
formulation the total derivative term should be omitted while
from the Lagrangian point of view it can be compensated
by adding to the bulk action a proper boundary term.

In this note we demonstrate how this general approach works
in the case of antisymmetric tensor fields of the second rank
and compute the two-point function of the corresponding local CFT
operators.

We start with the following action for  a massive complex
antisymmetric tensor field of the second rank\footnote{
It follows from \cite{KWN,GM} that antisymmetric tensor fields arising in
$S^5$ compactification of $IIB$ supergravity are classified by complex
representations of $SO(6)$.}
\bea
\la{act}
S=-\int d^5x \l \frac{i}{2}
\eps^{\mu\nu\rho\lambda\s}a^*_{\mu\nu}\p_{\rho}a_{\lambda\s}
+\sqrt{-g} m a^*_{\mu\nu} a^{\mu\nu} \r .
\eea
Here $g=-x_0^{-10}$ is the determinant of the $AdS$ metric:
$ds^2=\frac{1}{x_0^2}(dx_0^2+\eta_{ij}dx^idx^j)$. Because of infrared
divergencies one should regularize the action by cutting $AdS_5$ space
off at $x_0=\eps$ and leaving the part $x_0\geq \eps$. We use
the convention $\eps_{01234}=-\eps^{01234}=1$.

Action (\ref{act}) vanishes on shell and, therefore, can not produce
the two-point functions in the boundary CFT. According to the general scheme
discussed above, we need to rewrite action (\ref{act}) in
the form suitable for passing to the Hamiltonian
formulation. To this end one has to establish a proper set of variables
that can be treated as coordinates and their conjugate momenta. It can be
done by studying solutions of equations of motion coming from (\ref{act})
\bea
\la{eq}
\frac{i}{2}
\eps^{\mu\nu\rho\lambda\s}\p_{\rho}a_{\lambda\s}
+\sqrt{-g} m a^{\mu\nu}=0.
\eea
Acting on (\ref{eq}) with an operator
$-\frac{i}{2}\eps_{\mu\nu\rho\lambda\s}\n^{\rho}
+\sqrt{-g} m g_{\mu\lambda}g_{\nu\s}$
we arrive at the second-order equation
\bea
\la{seq}
\n^{\rho}\l \n_{\rho}a_{\mu\nu}-\n_{\mu}a_{\rho\nu}+\n_{\nu}a_{\rho\mu} \r
-m^2a_{\mu\nu}=0.
\eea
The last equation implies the constraint $\n^{\mu}a_{\mu\nu}=0$ and,
therefore, can be written in the form
\bea
\la{eq1}
\n^{\rho}\n_{\rho}a_{\mu\nu}+(6-m^2)a_{\mu\nu}=0.
\eea
Specifying (\ref{eq1}) for $a_{ij}$, we obtain
\bea
\la{aij}
x_0^2\p_0^2a_{ij} + x_0\p_0a_{ij} +x_0^2\Box a_{ij}
-m^2a_{ij}-2x_0(\p_ia_{0j}-\p_ja_{0i})=0,
\eea
where $\Box=\eta^{ij}\p_i\p_j$. The derivatives $\p_ia_{0j}$ can be
expressed from (\ref{eq}):
\bea
\la{der}
\p_ia_{0j}-\p_ja_{0i} = \p_0 a_{ij} + \frac{i}{2}x_0^{-1}m \eps_{ijkl}a^{kl},
\eea
where in the last formula and below the indices are raised with
respect to the Minkowski metric.
Therefore, eq.(\ref{aij}) reduces to
\bea
\la{aij1}
x_0^2\p_0^2a_{ij} - x_0\p_0a_{ij} +x_0^2\Box a_{ij}
-m^2a_{ij} - i m \eps_{ijkl}a^{kl}=0,
\eea
To solve (\ref{aij1}) we introduce the projections $a^{\pm}_{ij}$ on the
(anti)self-dual parts of $a_{ij}$:
\bea
\la{sd}
a^{\pm}_{ij}=\frac{1}{2}\left(a_{ij}\pm\frac{i}{2}\eps_{ijkl}a^{kl}\right),~~~~
a^{\pm}_{ij}=\pm \frac{i}{2}\eps_{ijkl}a^{\pm kl}.
\eea
Then eq.(\ref{aij1}) splits into equations for $a^+_{ij}$ and $a^-_{ij}$:
\bea
\la{apm}
x_0^2\p_0^2a^{\pm}_{ij} -x_0\p_0a^{\pm}_{ij} -m(m\pm 2)a^{\pm}_{ij}+x_0^2\Box a^{\pm}_{ij}=0.
\eea
Momentum space solutions of (\ref{apm}) for $\vec{k}^2>0$
obeying the boundary conditions
$a^{\pm}_{ij}(\eps,\vec{k})=a^{\pm}_{ij}(\vec{k})$
and vanishing at $x_0=\infty$ read
as\footnote{
As was noted in \cite{BKL} for $\vec{k}^2<0$ there are two independent solutions
regular in the interior. However, both of them are nonvanishing at infinity.
A proper account of these solutions may be achieved by introducing an
additional boundary at $x_0=1/\eps$ and requiring the vanishing of the
solution on this boundary. Then the solution is unique and
in the limit $\eps\to 0$ delivers the same contribution to the two-point
function as the solution for $\vec{k}^2>0$ does.
In the sequal, we restrict ourselves to the case $\vec{k}^2>0$.
}
\bea
\la{sol}
a^{\pm}_{ij}(x_0,\vec{k})=
\frac{x_0}{\eps}\frac{K_{m\pm 1}(x_0 k)}
{K_{m\pm 1}(\eps k)}a^{\pm}_{ij}(\vec{k}),
\eea
where $K_{m\pm 1}$ is the Mackdonald function and $k=|\vec{k}|$.

However, we can not assign arbitrary boundary values for both $a^{+}_{ij}(\vec{k})$
and $a^{-}_{ij}(\vec{k})$ since these components are related to each other.
To find this relation, note that the components $a_{0i}$ can be directly
found from (\ref{eq}):
\bea
\la{a0i}
a_{0i}=
-\frac{i}{2m}x_0\eps_{ijkl}\p^ja^{kl}.
\eea
Then substituting into (\ref{der}) one obtains the constraint
\bea
\frac{x_0}{m}\l\p_j\p^k(a^+_{ik}-a^-_{ik})-\p_i\p^k(a^+_{jk}-a^-_{jk})\r=
\p_0 a_{ij} + \frac{m}{x_0}(a^+_{ij}-a^-_{ij}),
\eea
which after projecting on its (anti)self-dual part results into the
following equations
\bea
\la{con}
\pm\frac{x_0}{2m}\l
\Box a^{\pm}_{ij}+\Box a^{\mp}_{ij}+
2(\p_i\p^ka^{\mp}_{jk}-\p_j\p^ka^{\mp}_{ik})\r=
\p_0 a^{\pm}_{ij} \pm \frac{m}{x_0} a^{\pm}_{ij}.
\eea

With the solution for $a_{ij}$ at hand one can compute the derivative
$\p_0a_{ij}$. By using the following properties of the Mackdonald 
function
$$
K_{\nu+1}(z)-K_{\nu-1}(z)=\frac{2\nu}{z}K_{\nu},~~~
K_{\nu+1}(z)+K_{\nu-1}(z)=-2K_{\nu}'(z)
$$
one finds
\bea
\la{d}
&&\p_0 a^+_{ij}(x_0,\vec{k})=-\l\frac{m}{x_0}+\frac{k^2x_0}{2m}\r a^+_{ij}(x_0,\vec{k})
+\frac{k^2x_0^2}{2m\eps}\frac{K_{m-1}(x_0 k)}
{K_{m+1}(\eps k)}a^+_{ij}(\vec{k}), \\
\nonumber
&&\p_0 a^-_{ij}(x_0,\vec{k})=\l\frac{m}{x_0}+\frac{k^2x_0}{2m}\r a^-_{ij}(x_0,\vec{k})
-\frac{k^2x_0^2}{2m\eps}\frac{K_{m+1}(x_0 k)}
{K_{m-1}(\eps k)}a^-_{ij}(\vec{k}).
\eea

Finally, substituting eqs.(\ref{sol}) and (\ref{d}) into (\ref{con})
we find the relation between $a^-_{ij}(\vec{k})$ and $a^+_{ij}(\vec{k})$:
\bea
\la{scol}
a^-_{ij}(\vec{k})=-\frac{K_{m-1}(\eps k)}{K_{m+1}(\eps k)}
\l a^+_{ij}(\vec{k})+2\frac{(k_i a^+_{jl}(\vec{k})-k_j 
a^+_{il}(\vec{k}))k^l}{k^2} \r.  
\eea 
In the sequal, we restrict 
ourselves to the case $m>0$. When $\eps\to 0$ the ratio 
$\frac{K_{m-1}(\eps k)}{K_{m+1}(\eps k)}$ behaves as $(\eps k)^2$ 
for $m\geq 1$ and as $(\eps k)^{2m}$ for $0<m<1$. 
Thus, if we keep $a^{+}_{ij}$ finite in the limit 
$\eps\to 0$, then $a^{-}_{ij}$ tends to zero. Otherwise, keeping of 
$a^{-}_{ij}$ finite leads to divergency of $a^{+}_{ij}$. Therefore, 
only the $a^{+}_{ij}$ component can couple on the boundary with the 
CFT operator ${\cal O}_{ij}$. This conclusion can be also verified by 
considering the conformal transformations of $a_{ij}^+$ on the 
boundary induced by isometries of $AdS$.

Denote by $\xi^a$ a Killing vector of the background metric.
Under diffeomorphisms generated by $\xi$ the antisymmetric tensor
$a_{ij}$ transforms as follows
\bea
\d a_{ij}=\xi^{\rho}\p_{\rho} a_{ij}+a_{i\rho}\p_j\xi^{\rho}-a_{j\rho}\p_i\xi^{\rho}
\la{sym}
\eea
Note that the Killing vectors of the $AdS$ background can
be written as
\bea
\la{Kil}
&&\xi^0=x_0(A_kx^k+D),\\
\nonumber
&&\xi^i=-\frac{x_0^2-\eps^2}{2}A^i+\left(
-\frac{1}{2}(A^i x^2 -2 x^i A_k x^k)+Dx^i+\Lambda_j^i x^j+P^i
\right),
\eea
where $A^i,D,\Lambda_j^i,P^i$ generate on the boundary special
conformal transformations, dilatations, Lorentz transformations and shifts
respectively. Since $\p_i \xi^0\sim x_0$ and $a_{0i}, a^-_{ij}$ tend to zero
when $\eps\to 0$, in this limit one finds the following transformation law
for the boundary value of $a_{ij}^{+}$:
\bea
\nonumber
\d a_{ij}^{+}=\xi^k\p_k a_{ij}^{+} +\xi^0\p_0 a_{ij}^{+}
+\frac{1}{2}\l
a_{ik}^{+}(\p_j\xi^k-\p^k\xi_j)-a_{jk}^{+}(\p_i\xi^k-\p^k\xi_i)+
\p_k\xi^k a_{ij}^{+}  \r.
\eea
Recalling that $\p_0a_{ij}^{+}=-\frac{m}{x_0}a_{ij}^{+}+O(1)$
and taking into account the explicit form of the Killing vectors
we finally arrive at
\bea
\la{tran}
\d a_{ij}^{+}=\xi^k\p_k a_{ij}^{+} +(2-m)(A_kx^k+D) a_{ij}^{+}
+a_{ik}^{+}{\bf \Lambda}^k_{~j}-a_{jk}^{+}{\bf \Lambda}^k_{~i},
\eea
where ${\bf \Lambda}^{ij}=\Lambda^{ij}+x^iA^j-x^jA^i$. Eq.(\ref{tran})
is nothing but the standard transformation law for an antisymmetric tensor
with the conformal weight $2-m$ under the conformal mappings. Thus, on the
boundary $a_{ij}^{+}$ couples to the operator of conformal dimension
$\Delta=2+m$. In particular, for $m=1$ the antisymmetric tensor field
$a^+_{ij}$ transforms in ${\bf 6}_c$ irrep of $SU(4)$ and couples
on the boundary to the following YM operator \cite{FFZ}:
$$
{\cal O}^{AB}_{ij}=\bar{\psi}^A\sigma_{ij}\bar{\psi}^B+2i\phi^{AB}F^+_{ij}
$$
that obviously has the conformal weight 3.

It is clear from the discussion above that $a^+_{ij}$ plays the role of the
coordinate. Now rewriting action (\ref{act}) in the form
$\int(p\dot{q}-H(p,q))$ we get
\bea
\nonumber
S&=&-\int d^5x \l (a^-_{ij})^*\p_0 a^+_{ij}+\p_0(a^+_{ij})^* a^-_{ij}
+i \eps^{ijkl}(a^*_{0i}\p_ja_{kl}-a^*_{ij}\p_ka_{0l})
+\frac{m}{x_0}(a^*_{ij}a^{ij}+2a^*_{0i}a^{0i}) \r \\
\la{act1}
&+&\int d^5x\p_0\l (a^+_{ij})^* a^{-ij}\r.
\eea
The last term in (\ref{act1}) is a total derivative, which is
omitted in passing to the Hamiltonian formulation. Thus, the action
one should use in computing the Green functions is given by
\bea
\nonumber
{\bf S}=-\int d^5x \l (a^-_{ij})^*\p_0 a^+_{ij}+\p_0(a^+_{ij})^* a^-_{ij}
+i \eps^{ijkl}(a^*_{0i}\p_ja_{kl}-a^*_{ij}\p_ka_{0l})
+\frac{m}{x_0}(a^*_{ij}a^{ij}+2a^*_{0i}a^{0i}) \r .
\eea
In the Lagrangian
picture the total derivative term can be compensated by adding to
action (\ref{act1}) the following boundary term
\bea
\la{boun}
I=\int d^4x  (a^+_{ij})^* a^{-ij}.
\eea
Thus, the on-shell value of ${\bf S}$ is given by
\bea
\la{ons}
{\bf S}=-\int d^4k   \frac{K_{m-1}(\eps k)}{K_{m+1}(\eps k)}
(a^{+ij})^*\l a^+_{ij} + 2\frac{(k_ia^+_{jl}-k_ja^+_{il})k^l}{k^2}\r.
\eea
When $\eps\to 0$ and for $m$ integer one finds
$$
\frac{K_{m-1}(\eps k)}{K_{m+1}(\eps k)}=
\frac{(-1)^m}{2^{2m-1}(m-1)!m!}(\eps k)^{2m}\log{\eps k}+...,
$$
while for non-integer $m$:
$$
\frac{K_{m-1}(\eps k)}{K_{m+1}(\eps k)}=
-\frac{\Gamma(2-m)}{2^{2m}(m-1)\Gamma(m+1)}(\eps k)^{2m}+...,
$$
where in both cases we indicated only the first non-analytical term.
Hence, from (\ref{ons}) we deduce the two-point function of ${\cal O}$
in the boundary CFT:
\bea
\nonumber
\langle \bar{{\cal O}}^{ij}(\vec{k}){\cal O}^{kl}(\vec{q})\rangle =
-\d(\vec{k}+\vec{q})
\frac{(-1)^m}{2^{2m-2}(m-1)!m!}(\eps k)^{2m}\log{\eps k}
\l
\eta^{i[k}\eta^{l]j} + 2\frac{(k^i\eta^{j[k}-k^j\eta^{i[k})k^{l]}}{k^2}
\r
\eea
and a similar result for $m$ non-integer. The last expression 
exhibits the structure of the correlation function for an antisymmetric 
tensor field of the conformal weight $2+m$ in the $D=4$, 
${\cal N}=4$ SYM theory.

Note that on shell instead of (\ref{boun}) one can use the following
boundary term
\bea
\nonumber
I=\frac{1}{2}\int d^4x  a_{ij}^* a^{ij}.
\eea
Finally, we remark that in the case $m<0$ the component $a_{ij}^-$
should be regarded as the coordinate that leads to the change of the sign
in the last formula.

\vskip 1cm

{\bf ACKNOWLEDGMENT} The authors thank L.O.Chekhov
for valuable discussions. This work has been supported in part
by the RFBI grants N96-01-00608 and N96-01-00551.

\newpage

\end{document}